\documentclass{iopart}

\usepackage{graphicx}
\usepackage{cite}
\expandafter\let\csname equation*\endcsname\relax
\expandafter\let\csname endequation*\endcsname\relax
\usepackage{amsmath}
\usepackage{amssymb}

\usepackage{xcolor}
\begin{document}

   \title{Thermodynamic Coefficients of Ideal Fermi--gas}


\author{A S Kozharin and P R Levashov}

\address{Joint Institute for High Temperatures RAS, Izhorskaya 13 Bldg 2, Moscow 125412, Russia}
\address{Moscow Institute of Physics and Technology, 9 Institutskiy per., Dolgoprudny, Moscow Region, 141700, Russia}
\ead{1alekseik1@gmail.com}
\vspace{10pt}
\begin{indented}
\item[]February 2020
\end{indented}

\begin{abstract}
  We present analytical formulae for the first and second derivatives of the Helmholtz free energy of non-relativistic ideal Fermi gas. Important thermodynamic quantities such as heat capacity, sound velocity, heat capacity ratio and others can be easily expressed through the derivatives. We demonstrate correct ideal Boltzmann gas and low--temperature Fermi--gas asymptotes and derive corrections to thermodynamic functions for these limiting cases. Numerical computations of thermodynamic properties of  ideal Fermi--gas can be accurately performed using the developed freely available Python module \texttt{ifg}.
\end{abstract}


%


\section{Introduction}
  The model of non-interacting fermions or the ideal Fermi--gas (IFG) model \cite{Fermi:ZP:1926} is widely used in atomic physics, astrophysics and condensed matter physics. The model was formulated almost a hundred years ago immediately after the formulation of the famous Pauli exclusion principle \cite{Pauli:ZP:1925} and Fermi--Dirac statistics \cite{Fermi:ZP:1926,Dirac:PRSL:1926}. The appearance of the IFG model served as the basis of quantum statistical physics and made it possible to explain a lot of peculiar physical phenomena, in particular, evolution and collapse of stars \cite{Chandrasekhar:PM:1931}, nova explosions, properties of semiconductors and metals \cite{Ashcroft:book:2011}. Also the IFG model is widely used in semiempirical equations of state for the description of the electronic subsystem in matter \cite{Eliezer:EOSbook:2002,Fortov:EOSbook:2016}. As in metals at relatively low temperatures the electronic subsystem gives a major contribution to thermodynamic properties it is reasonable to approximate the shock Hugoniot of a metal by the IFG Hugoniot~\cite{Nellis:JAP:2003}. Numerous textbooks on statistical physics include a survey of thermodynamic properties of the IFG model \cite{Landau:statmech:1958,Balescu:book:1975,Huang:book:1987}. Usually the analytical expressions for pressure and internal energy are presented; the important asymptotic limits of the degenerate Fermi--gas at $T = 0$ and ideal Boltzmann gas (IBG) at very high temperatures are also discussed. For practical applications the low-temperature asymptotic formula for the internal energy of IFG is derived from which it follows the linear dependence of electronic isochoric heat capacity on temperature \cite{Landau:statmech:1958}. However, a consistent derivation of the so-called thermodynamic coefficients---thermodynamic quantities containing second derivatives of thermodynamic potential---of IFG still cannot be found in literature.

  In this paper we derive analytical expressions for the first and the second derivatives of the Helmholtz free energy of non-relativistic IFG. All important thermodynamic coefficients including isochoric and isobaric heat capacity, adiabatic sound velocity, heat capacity ratio and all others can be expressed through the derivatives. For practical calculations we use recent high-accuracy approximations for direct and inverse Fermi integrals \cite{Fukushima:AMC:2015}. A freely available Python module \texttt{ifg} \cite{ifg} was created to compute thermodynamic functions of IFG.

\section{IFG model}

%

   IFG is a system of many non-interacting fermions, i.e. particles with half odd integer spin. 
   According to the Pauli exclusion principle \cite{Pauli:ZP:1925,Pauli:PR:1940}, no quantum state can be occupied by more than one fermion with an identical set of quantum numbers.
   Therefore particles of IFG cannot all in one occupy the ground state at zero temperature.
   There are several consequences of this fact.
   Firstly, the total energy of IFG at absolute zero is higher than a sum of its particles' ground energies.
   Secondly, the pressure of IFG is not equal to zero even at zero temperature, in contrast to that of IBG.

   We use the atomic system of units throughout the paper in which the reduced Planck constant, electron mass and charge equal to unity. Additionally, the Boltzmann constant is also set to unity. 

\subsection{Chemical potential and its derivatives}

   For a system of fermions the Fermi-Dirac statistics should be considered \cite{Landau:statmech:1958}: the average number of fermions $n_k$ in the $k$-th quantum state is given by the formula:
   
\begin{equation}
  n_k = \frac{1}{\exp{[(\varepsilon_k - \mu)/T]} + 1}.
  \label{eq:fd}
\end{equation}
Here $\mu$ is chemical potential, $T$ is the temperature of the system, $\varepsilon_k$ is the energy of the $k$th state. To obtain the total number of fermions $N$ we should take into account the spin degeneracy $g = 2s + 1$, the absence of interaction and elementary volume of phase space $(2\pi)^3$. Here $s$ is spin of each fermion. Thus it may be shown \cite{Landau:statmech:1958} that
   \begin{equation}
       \label{eq:conc}
        \frac{N}{V}=\frac{gm_r^{3/2}}{\sqrt{2} \pi^{2}} \int_{0}^{\infty} \frac{\sqrt{\varepsilon} d \varepsilon}{e^{(\varepsilon-\mu) / T}+1}.
   \end{equation}
Here $V$ is the volume of the system, $m_r$ is the fermion mass relative to the electron mass. Defining $\bar g = gm_r^{3/2}$, dimensionless chemical potential $y = \mu/T$ and volume per one particle $v = V/N$ expression~(\ref{eq:conc}) can be rewritten as follows:
   \begin{equation}
      \frac{1}{v} = \frac{\bar g T^{3/2}}{\sqrt{2}\pi^2}I_{1/2}(y).
       \label{eq:mu_equation}
   \end{equation}
Here the so-called Fermi-Dirac function is introduced:
   \begin{equation}
       \label{eq:fermi-dirak_integral_definition}
    I_{j}(y)= \int_{0}^{\infty} \frac{t^{j}}{e^{t-y}+1} dt
   \end{equation}
Equation~(\ref{eq:mu_equation}) is a parametric function of $\mu$ and may be inversed to find $\mu$ at given $T$ and $v$.

To find all first and second derivatives of $\mu (T, v)$ we need to differentiate (\ref{eq:mu_equation}) with respect to corresponding variables and then simplify the results using (\ref{eq:mu_equation}).

The derivatives of $y = \mu /T$ with respect to temperature and volume per particle are:
 
   \begin{equation}
       \label{eq:mu_T}
       y'_{T} = -\frac{3 I_{1 / 2}(y)}{T I_{-1 / 2}(y)},
   \end{equation}

   \begin{equation}
       \label{eq:mu_v}
       y'_{v} = - \frac{2 I_{1 / 2} (y)}{v I_{-1 / 2} (y)},
   \end{equation}

   \begin{equation}
       \label{eq:mu_vv}
       y''_{vv} = \frac{4 I_{1 / 2} (y)}{v^2 I_{-1 / 2} (y)}\, + \frac{2 I_{-3 /2} (y) I_{1 / 2} ^2 (y)}{v^2 I_{-1 /2}^3 (y)}, 
   \end{equation}

   \begin{equation}
       \label{eq:mu_TT}
       y''_{TT} = \frac{15 I_{1 /2} (y)}{2 T^2 I_{-1 /2} (y)}\, + \frac{9 I_{-3 /2}(y) I_{1 /2}^2 (y)}{2 T^2 I_{-1 /2}^3 (y)}, 
   \end{equation}

   \begin{equation}
       \label{eq:mu_vT}
       y''_{vT} = y''_{Tv} = \frac{3 I_{1 /2}(y)}{T v I_{-1 /2} (y)}\, + \frac{3 I_{-3 /2} (y) I_{1 /2}^2 (y)}{T v I_{-1 /2}^3 (y)}.
   \end{equation}

\subsection{Helmholtz free energy and thermodynamic functions}

The Helmholtz free energy of IFG per particle has the following expression \cite{kirzhnits:UFN:1975}:
   \begin{equation}
       \label{eq:helmholtz_potential}
       F = \frac{\bar g}{\sqrt{2}\pi^2}\,  T^{5 / 2} v \left( y \, I_{1 /2} (y) - \frac{2}{3}\, I_{3 / 2} (y) \right) = T\left[
        y - \frac{2}{3}\frac{I_{3/2}(y)}{I_{1/2}(y)}
       \right].
   \end{equation}

The last expression is obtained by substituting $v$ given by (\ref{eq:mu_equation}) into (\ref{eq:helmholtz_potential}). All thermodynamic properties of IFG can now be expressed through the derivatives of $F$:

  \begin{equation}
    \label{eq:PES}
    P = -F'_v;\quad E = F - TF'_T;\quad S = -F'_T;
  \end{equation}
  \begin{equation}
    \label{eq:CVCP}
    C_V = -TF''_{TT};\quad C_P = -TF''_{TT} + \frac{T(F''_{vT})^2}{F''_{vv}};
  \end{equation}
  \begin{equation}
    \label{eq:CTCS}
    C_T^2 = v^2F''_{vv};\quad C_S^2 = v^2F''_{vv} - \frac{v^2(F''_{vT})^2}{F''_{TT}};\quad \gamma = \frac{VF''_{vT}}{TF''_{TT}};\ \ldots
  \end{equation}

Here $P$ is pressure, $C_V$ and $C_P$ are isochoric and isobaric heat capacities, respectively, $C_T$ and $C_S$ are isothermal and adiabatic sound velocities, respectively, $\gamma$ is Gr\"uneisen parameter. 

With the help of expressions (\ref{eq:mu_T})--(\ref{eq:mu_vT}) the first and second derivatives of $F$ with respect to $v$ and $T$ can be represented as follows:

\begin{equation}
  \label{eq:F_vF_T}
  F'_v = -\frac{\bar g\sqrt{2}T^{5/2}I_{3/2}(y)}{3\pi^2},\quad 
  F'_T = \frac{\bar g\sqrt{2}vT^{3/2}[3yI_{1/2}(y) - 5I_{3/2}(y)]}{6\pi^2},
\end{equation}

\begin{equation}
  \label{eq:F_vv}
  F''_{vv} = \frac{\bar g\sqrt{2}T^{5/2}I_{1/2}^2(y)}{\pi^2 v I_{-1/2}(y)},
\end{equation}

\begin{equation}
  \label{eq:F_TT}
  F''_{TT} = -\frac{\bar g\sqrt{2}vT^{1/2}[5I_{-1/2}(y)I_{3/2}(y) - 9I_{1/2}^2(y)]}{4\pi^2I_{-1/2}(y)},
\end{equation}

\begin{equation}
  \label{eq:F_vT}
  F''_{vT} = F''_{Tv} = -\frac{\bar g\sqrt{2}T^{3/2}[5I_{-1/2}(y)I_{3/2}(y) - 9I_{1/2}^2(y)]}{6\pi^2I_{-1/2}(y)}.
\end{equation}

Using formulae (\ref{eq:PES})--(\ref{eq:CTCS}), (\ref{eq:F_vF_T})--(\ref{eq:F_vT}) and (\ref{eq:mu_equation}) thermodynamic properties of IFG can be expressed through Fermi--Dirac functions:

   \begin{equation}
       \label{eq:pressure}
       P = \frac{\bar g \sqrt{2} T^{5 /2} I_{3 / 2}(y)}{3 \pi^{2}} = \frac{2T}{3v}\frac{I_{3/2}(y)}{I_{1/2}(y)},
   \end{equation}

   \begin{equation}
       \label{eq:energy}
       E = \frac{\bar g v}{\sqrt{2}\pi^2} T^{5 /2} I_{3 / 2}(y) = \frac{I_{3/2}(y)}{I_{1/2}(y)}T,
   \end{equation}

   \begin{equation}
       \label{eq:entropy}
       S = -\frac{\bar g\sqrt{2} T^{3 /2} v\left(3 I_{1 / 2}(y) y-5 I_{3 / 2}(y)\right)}{6 \pi^{2}} = -\frac{3I_{1/2}(y) - 5I_{3/2}(y)}{3I_{1/2}(y)},
   \end{equation}

   \begin{equation}
       \label{eq:C_v}
       C_v = \frac{\bar g\sqrt{2} T^{3 /2} v \left( 5 I_{-1 / 2 } (y) I_{3 /2} (y) - 9 I_{1 / 2}^2 (y)\right)}{4\pi^2 I_{-1 / 2} (y)}\ = 
       \frac{5}{2}\,\frac{I_{3/2}(y)}{I_{1/2}(y)} - \frac{9}{2}\,\frac{I_{1/2}(y)}{I_{-1/2}(y)}, 
   \end{equation}

   \begin{multline}
       \label{eq:C_P}
          C_{P}= 
            \frac{5 \bar g \sqrt{2} T^{3 / 2} v\left(5 I_{-1 / 2}(y) I_{3 / 2}(y)-9 I_{1 / 2}^{2}(y)\right) I_{3 / 2}(y)}{36 \pi^{2} I_{1 / 2}^{2}(y)} = {}\\
            \frac{25}{18}\,\frac{I^2_{3/2}(y)I_{-1/2}(y)}{I^3_{1/2}(y)} -
            \frac{5}{2}\,\frac{I_{3/2}(y)}{I_{1/2}(y)},
   \end{multline}

   \begin{equation}
       \label{eq:C_T}
       C_{T}^2 = \cfrac{\sqrt{2}\bar g}{\pi^2}\,  \cfrac{T^{5 /2} v I_{1 / 2}^2 (y)}{I_{-1 / 2} (y)} = \frac{2TI_{1/2}(y)}{I_{-1/2}(y)},
   \end{equation}

   \begin{equation}
       \label{eq:C_S}
       C_{S}^2 =  
       \cfrac{5 \sqrt{2}\bar g}{9 \pi^2}\, T^{5 / 2} v I_{3 / 2} (y) = \frac{10T I_{3/2}(y)}{9I_{1/2}(y)}.
   \end{equation}

   It is interesting to note that the last two formulae for isothermal and isentropic sound speed look remarkably simple. Also Gr\"uneisen parameter for IFG is always $2/3$ as it can be easily seen from (\ref{eq:CTCS}), (\ref{eq:F_TT}) and (\ref{eq:F_vT}).

   \subsection{Asymptotic formulae at high temperatures}

   At high temperatures IFG becomes ordinary IBG.
   Thus, all thermodynamic properties of IFG should tend to their classical limits at $y = \mu/T \ll -1$.
   To demonstrate this, two notices should be made.

   Firstly, at $y\ll -1$ we can neglect unity in the denominator of (\ref{eq:fermi-dirak_integral_definition}) and obtain 
   
   \begin{equation}
       \label{eq:asymp-fermi-dirac}
       I_{k}(y)\approx \Gamma(k + 1)e^{y},
   \end{equation}
   in particular,
   $$
    I_{-1/2}(y)\approx \sqrt{\pi}e^{y};\quad I_{1/2}(y)\approx \frac{\sqrt{\pi}}{2}e^{y};\quad
    I_{3/2}(y)\approx \frac{3\sqrt{\pi}}{4}e^{y}.
   $$

   Secondly, the left hand side of (\ref{eq:mu_equation}) should be finite at $T \rightarrow \infty$, therefore $I_{1/2}(y)$ should tend to zero or $\mu/T\rightarrow -\infty$. That means that $\mu\rightarrow -\infty$. Indeed, at $y\ll -1$ chemical potential derived from (\ref{eq:mu_equation}) becomes classical:
   \begin{equation}
    \mu = \mu_B = T\ln\left[
      \frac{1}{\bar gv}\left(
        \frac{2\pi}{T}
      \right)^{3/2}
    \right],
   \end{equation}
and obviously goes to $-\infty$ at $T \rightarrow \infty$.

   Using (\ref{eq:asymp-fermi-dirac}) the following asymptotic expressions for IFG thermodynamic functions may be obtained from (\ref{eq:helmholtz_potential}) and (\ref{eq:pressure})--(\ref{eq:C_S}) at $y\ll -1$:
   \begin{equation}
        F\approx (y - 1)T = \mu - T = -T\ln\left[
          \bar gev\left(
            \frac{T}{2\pi}
          \right)^{3/2}
        \right],\quad  
   \end{equation}
    \begin{equation}
        P\approx \cfrac{T}{v},\quad  
        E\approx \cfrac{3}{2}T,\quad  
        S\approx \cfrac{5}{2} - y,
    \end{equation}

    \begin{equation}
        C_{V}\approx \cfrac{3}{2},\quad
        C_{P}\approx \cfrac{5}{2},         
    \end{equation}

    \begin{equation}
        C_{T}^2\approx T,\quad
        C_{S}^2\approx \cfrac{5}{3} T.
    \end{equation}

The first correction to (\ref{eq:asymp-fermi-dirac}) can be found using the series expansion of the Fermi--Dirac function (\ref{eq:fermi-dirak_integral_definition}) by $\exp(y)\ll 1$:
\begin{equation}
  \label{eq:IhighT}
  I_j(y)\approx e^y \Gamma(j + 1) - \frac{e^{2y}}{2^{j+1}}\Gamma(j + 1),
\end{equation}
in particular,
$$
  I_{1/2}(y)\approx \frac{\sqrt{\pi}e^y}{2}\left(
    1 - \frac{1}{2^{3/2}}e^y
  \right),\ 
  I_{3/2}(y)\approx \frac{3\sqrt{\pi}}{4}\left(
    1 - \frac{1}{2^{5/2}}e^y
  \right).
$$
Substituting the asymptotic expression (\ref{eq:IhighT}) for $I_{1/2}$ into (\ref{eq:mu_equation}) and replacing $y$ in the second term by $\mu_B/T$ we obtain
\begin{equation}
  \label{eq:mu_equation_high_T}
  \frac{1}{\bar gv}\left(
    \frac{2\pi}{T}
    \right)^{3/2} = \exp\left(\frac{\mu_B}{T}\right)
    = e^y\left[
    1 - \frac{\pi^{3/2}}{\bar gvT^{3/2}}
  \right].
\end{equation}
As in expression (\ref{eq:mu_equation_high_T}) the second term in square brackets is small compared to unity, we may approximately solve (\ref{eq:mu_equation_high_T}) with respect to $y$:
\begin{equation}
  \label{eq:muhighT}
  e^y \approx e^{\mu_B / T}\left(
    1 + \frac{\pi^{3/2}}{\bar gvT^{3/2}}
  \right),
\end{equation}
or, within the same accuracy,
$$
  y = \frac{\mu_B}{T} + \frac{\pi^{3/2}}{\bar gvT^{3/2}}.
$$

Substituting the asymptotic expression (\ref{eq:IhighT}) for $I_{3/2}$ into (\ref{eq:pressure}) we obtain for pressure:
\begin{equation}
  P = \frac{\bar gT^{5/2}e^y}{(2\pi)^{3/2}}\left(
    1 - \frac{e^y}{2^{5/2}}
  \right).
\end{equation}

If we replace $e^y$ by expression (\ref{eq:muhighT}) and leave only the terms with $T$ and $T^{-1/2}$ the following formula may be derived \cite{Landau:statmech:1958}:
\begin{equation}
  \label{eq:p_ht}
  P \approx \frac{T}{v} + \frac{\pi^{3/2}}{2\sqrt{T}\bar gv^2}.
\end{equation}
Note the sign ``+'' before the second term that indicates additional repulsion between fermions due to degeneracy.

Now we can find the asymptotic expression for Helmholtz free energy per particle (again leaving only the terms with powers of $T$ greater or equal to $-1/2$):
\begin{equation}
  \label{eq:FhighT}
  F = \mu - Pv = (\mu_B - T) + \frac{\pi^{3/2}}{2\bar gvT^{1/2}}.
\end{equation}
Other asymptotic expressions can be found by differentiation of (\ref{eq:FhighT}) using (\ref{eq:PES})--(\ref{eq:CTCS}):

\begin{equation}
  \label{eq:EShighT}
  E = \frac{3T}{2} + \frac{3\pi^{3/2}}{4\sqrt{T}\bar gv};\quad S = \frac{5}{2} - \ln\left[
    \frac{1}{gv}\left(
      \frac{2\pi}{T}
    \right)^{3/2}
  \right] + \frac{\pi^{3/2}}{4T^{3/2}\bar gv};
\end{equation}
\begin{equation}
  \label{eq:CVCPhighT}
  C_V = \frac{3}{2} - \frac{3\pi^{3/2}}{8T^{3/2}\bar gv};\quad 
  C_P = \frac{5}{2} - \cfrac{15 \pi^{3 / 2}}{8 T^{3 / 2}\bar g v}\, ;
\end{equation}
\begin{equation}
  \label{eq:CTCShighT}
  C_T^2 = T + \frac{\pi^{3/2}}{\sqrt{T}\bar gv};\quad 
  C_S^2 = \frac{5 T}{3} + \frac{5 \pi^{3/2}}{6 \sqrt{T}\bar gv} .
\end{equation}
%
It can be clearly seen from expressions (\ref{eq:CVCPhighT}), (\ref{eq:CTCShighT}) that degeneracy effects reduce heat capacity and increase sound velocity.


\subsection{Asymptotic formulae at low temperatures}

Low temperatures correspond to the case $y = \mu/T \gg 1$. Under this condition, the following asymptotic series expansion for the Fermi--Dirac functions also known as a Sommerfeld expansion \cite{Sommerfeld:ZP:1928,Ashcroft:book:2011} is valid \cite{Shemyakin:JPA:2010}:
%
%
\begin{equation}
\label{eq:Ihalf}
I_{1/2}(y) \approx
\frac{2y^{3/2}}{3}\left[1+\frac{\pi^2}{8y^2}+\frac{7\pi^4}{640y^4} + \ldots\right],
\end{equation}
\begin{equation}
\label{eq:I3half}
I_{3/2}(y) \approx \frac{2y^{5/2}}{5}\left[1+\frac{5\pi^2}{8y^2}-\frac{7\pi^4}{384y^4} + \ldots\right].
\end{equation}

If we leave only the first term of the expansion (\ref{eq:Ihalf}) then expression (\ref{eq:mu_equation}) will be independent of temperature; it is equivalent to the case $T=0$ and chemical potential becomes equal to the well-known Fermi energy:
$$\mu|_{T = 0} = \varepsilon_F = \left(
  \frac{3\pi^2}{\sqrt{2}\bar gv}
\right)^{2/3}.
$$
Leaving the first two terms in (\ref{eq:Ihalf}) and substituting $y = \varepsilon_F / T$ only to the second term one may easily obtain the first correction to the chemical potential from (\ref{eq:mu_equation}):
\begin{equation}
  \mu \approx \varepsilon_F\left[
    1 - \frac{\pi^2}{12}\left(
      \frac{T}{\varepsilon_F}
    \right)^2
  \right].
  \label{eq:mulowtemp}
\end{equation}
Substituting (\ref{eq:Ihalf}) and (\ref{eq:I3half}) into (\ref{eq:helmholtz_potential}) and leaving only the first two terms we obtain:
\begin{equation}
  F \approx \frac{\bar g}{\sqrt{2}\pi^2}T^{5/2}v\left[
    \frac{2}{5}y^{5/2} - \frac{\pi^2}{12}y^{1/2}
  \right] = \frac{3T^{5/2}}{2\varepsilon_F^{3/2}}\left[
    \frac{2}{5}y^{5/2} - \frac{\pi^2}{12}y^{1/2}    
  \right].
  \label{eq:Flowtemp}
\end{equation}
Finally, we substitute (\ref{eq:mulowtemp}) into (\ref{eq:Flowtemp}) and leave only the first temperature correction. The asymptotic expression for $F$ now becomes:
\begin{equation}
  \label{eq:F_lt}
  F \approx \frac{3}{5}\varepsilon_F\left[
    1 - \frac{5\pi^2}{12}\left(
      \frac{T}{\varepsilon_F}
    \right)^2
  \right] = Av^{-2/3} - \frac{\beta}{2}T^2 v^{2/3}.
\end{equation}
Here $F_0$ is the free energy at $T = 0$, $\beta = (\bar g \pi / 6)^{2/3}$ is the so-called electron heat capacity coefficient \cite{Landau:statmech:1958},
$$
  A = \frac{3}{5}\left(
\frac{3\pi^2}{\sqrt{2}\bar g}
  \right)^{2/3}.
$$

Using formulae (\ref{eq:PES})--(\ref{eq:CTCS}) we obtain the following low-temperature expressions for thermodynamic functions:
\begin{equation}
  \label{eq:P_lt}
  P = \frac{2}{3}Av^{-5/3} + \frac{\beta}{3}T^2 v^{-1/3},
\end{equation}
$$
  E = Av^{-2/3} + \frac{\beta}{2}T^2 v^{2/3} = \frac{3}{5}\varepsilon_F + \frac{\beta}{2}T^2 v^{2/3},
$$
$$
  S = \beta T v^{2/3},
$$
\begin{equation}
  \label{eq:cv_lt}
  C_v = \beta T v^{2/3},
\end{equation}
$$
  C_P = \beta T v^{2/3},
$$
$$
  C_T^2 = \frac{10A}{9}v^{-2/3} + \frac{\beta}{9}T^2 v^{2/3},
$$
\begin{equation}
  \label{eq:cs2_lt}
  C_S^2 = \frac{10A}{9}v^{-2/3} + \frac{5\beta}{9}T^2 v^{2/3}.
\end{equation}
%

%


    \begin{figure}[!h]
      \includegraphics[width=0.5\columnwidth]{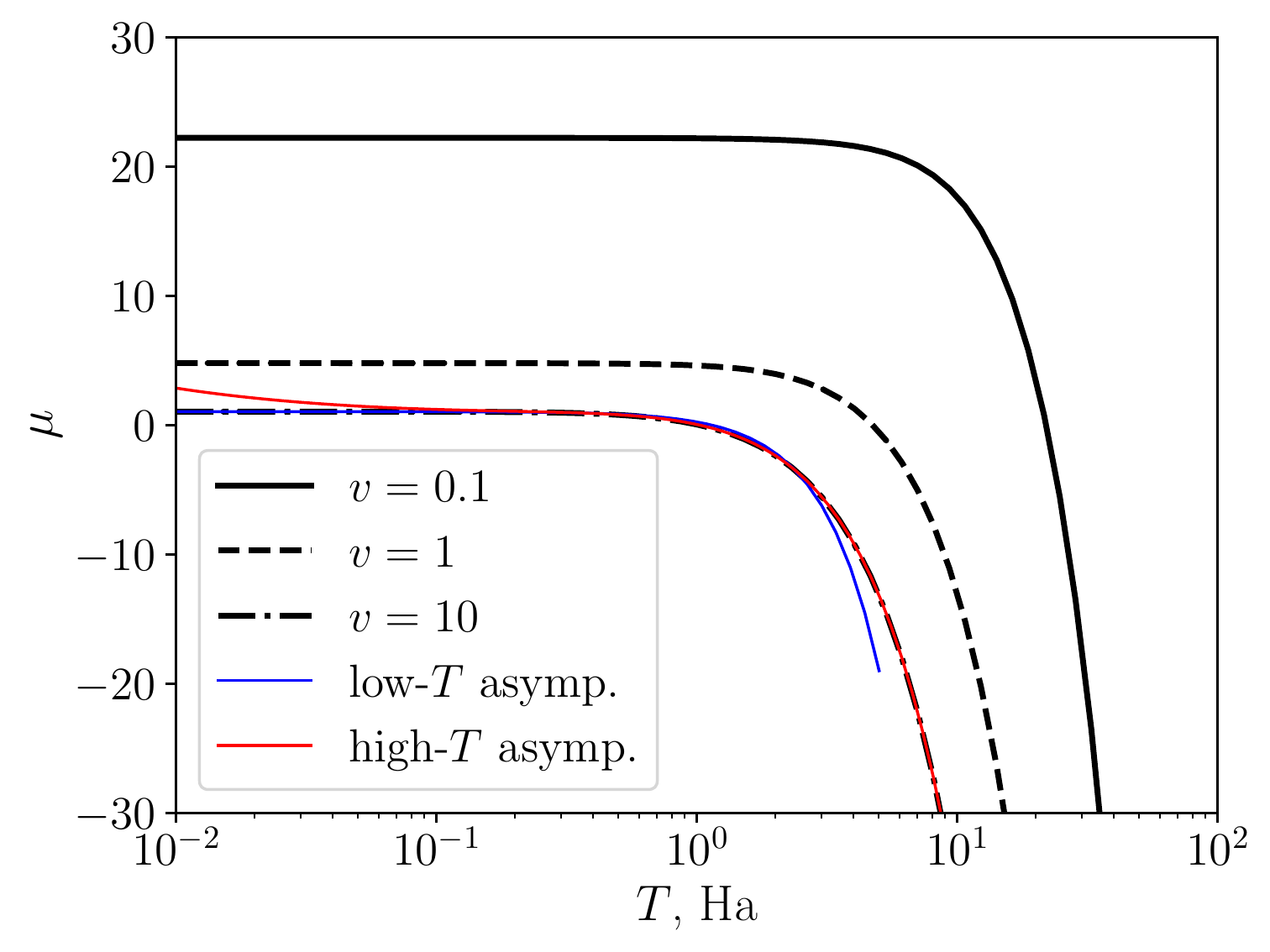}(a)
      \includegraphics[width=0.5\columnwidth]{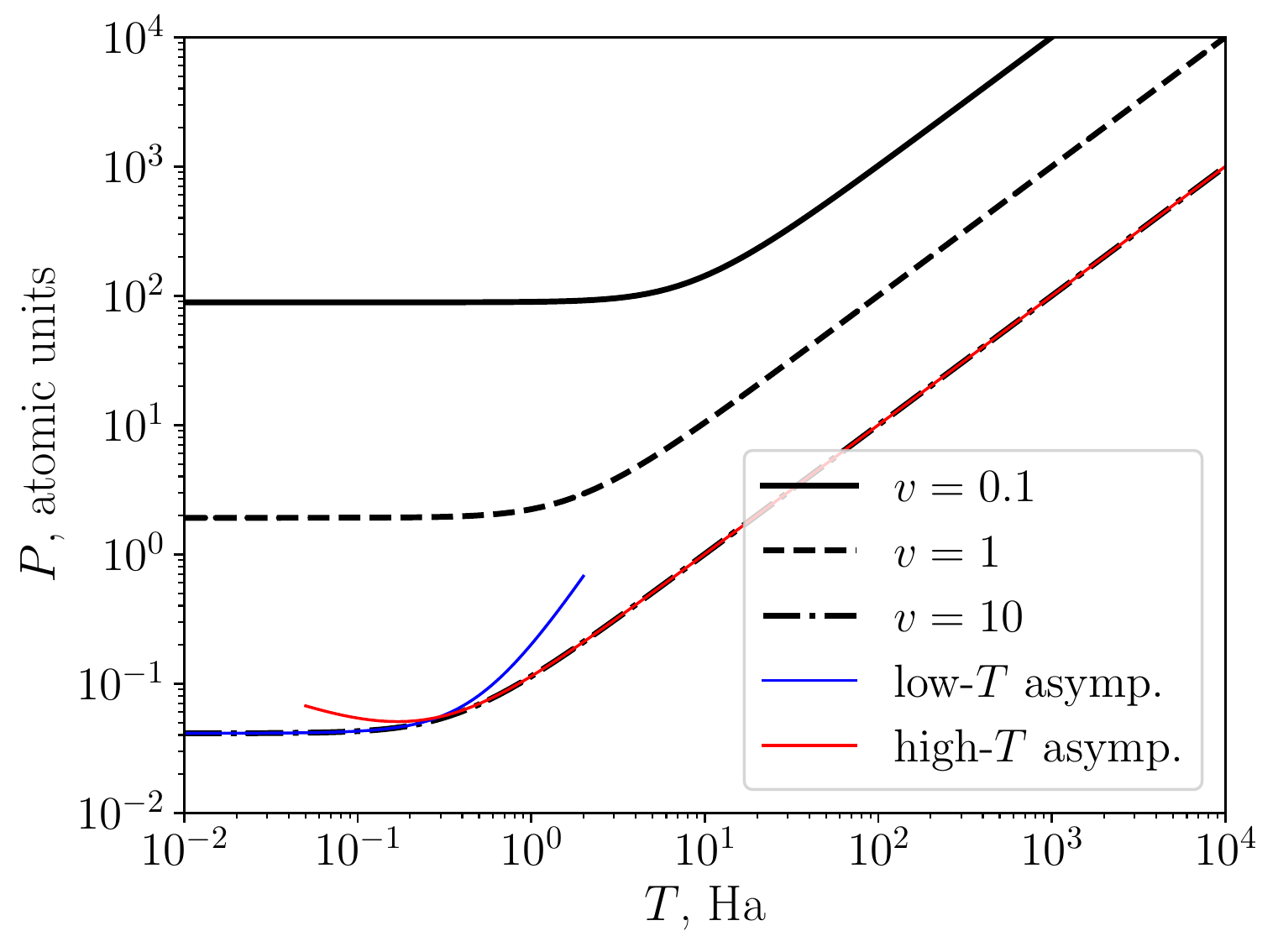}(b)
      \caption{Chemical potential (a) and pressure (b) of IFG at three isochores. Black lines: solid---$v = 0.1$, dashed---$v = 1$, dash-dotted---$v = 10$. Solid blue line---low-temperature asymptote (\ref{eq:mulowtemp}) for $\mu$, (\ref{eq:P_lt}) for $P$ at $v = 10$, solid red line---high-temperature asymptote (\ref{eq:muhighT}) for $\mu$, (\ref{eq:p_ht}) for $P$ at $v = 10$.}
      \label{fig:chemical_potential}
    \end{figure}

\subsection{Graphical illustration of IFG thermodynamic functions and their asymptotes}

    We apply the exact and asymptotic formulae obtained in the previous sections to electronic IFG with $g = 2$, $m_r = 1$. In figure~\ref{fig:chemical_potential} we illustrate the dependencies of chemical potential (a) and pressure (b) on temperature along three isochores. Chemical potential tends to $\varepsilon_F$ at $T\to 0$ and rapidly decreases and becomes negative at high temperatures. 

\begin{figure}[!h]
      \centering\includegraphics[width=0.75\columnwidth]{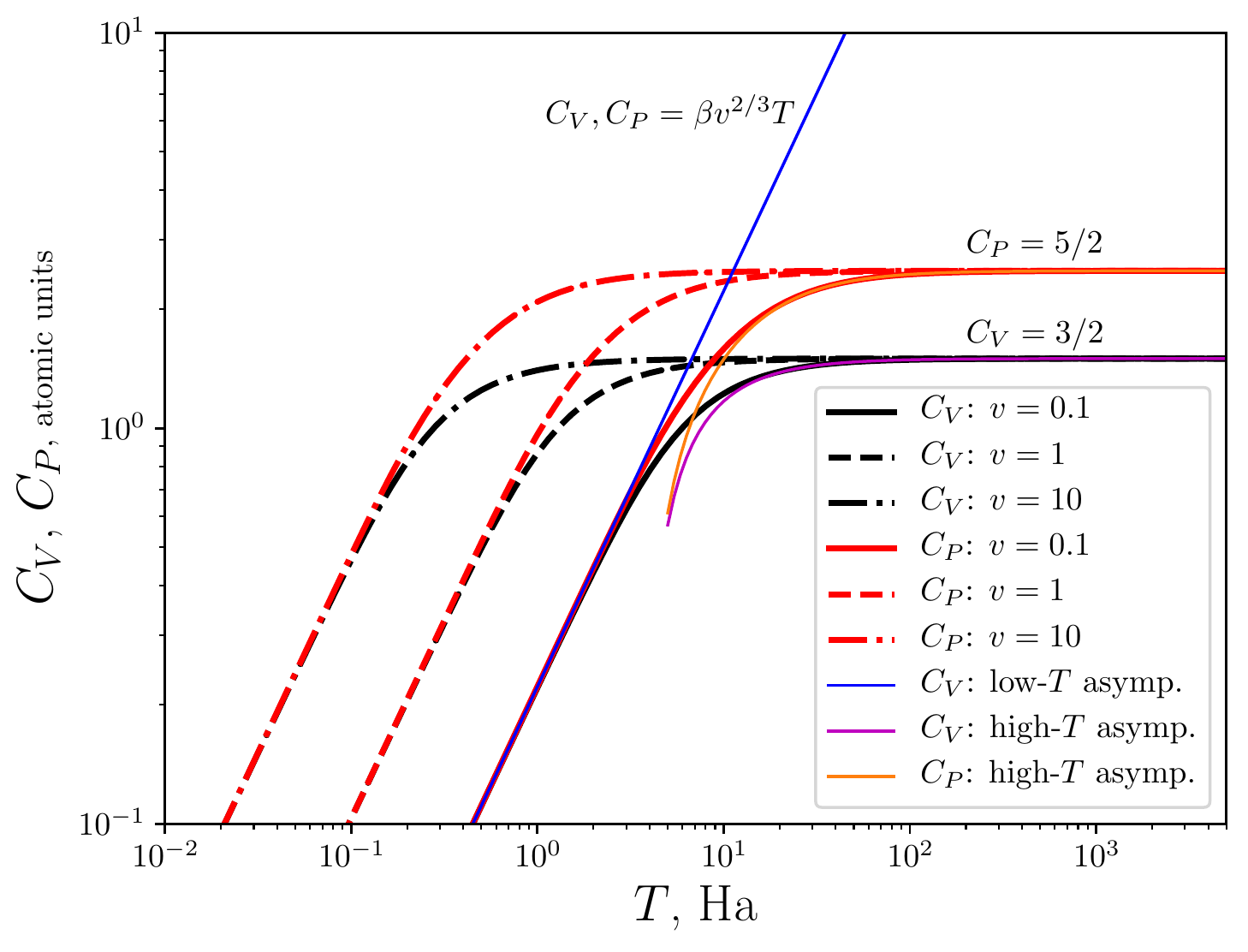}
      \caption{Isochoric and isobaric heat capacity of IFG at three isochores. Black lines---$C_V$, red lines---$C_P$. Solid lines---$v = 0.1$, dashed lines---$v = 1$, dash-dotted lines---$v = 10$. Solid blue line---low-temperature asymptote (\ref{eq:cv_lt}) for $C_V$ and $C_P$ at $v = 0.1$; solid magenta line---high-temperature asymptote (\ref{eq:CVCPhighT}) for $C_V$ at $v = 0.1$; solid organge line---high-temperature asymptote (\ref{eq:CVCPhighT}) for $C_P$ at $v = 0.1$.}
      \label{fig:heat_capacity}
    \end{figure}

    Isochoric and isobaric heat capacities vs. temperature are shown in figure~\ref{fig:heat_capacity}. At low temperatures both $C_V$ and $C_P$ demonstrate a linear temperature dependence. $C_P$ and $C_V$ can be distinguished at low temperatures only if we consider the terms proportional to $T^3$ \cite{Landau:statmech:1958} while the low-temperature approximation (\ref{eq:F_lt}) contains only the quadratic term. At high temperatures $C_V\to 3/2$ and $C_P\to 5/2$.

\begin{figure}[!h]
      \includegraphics[width=0.495\columnwidth]{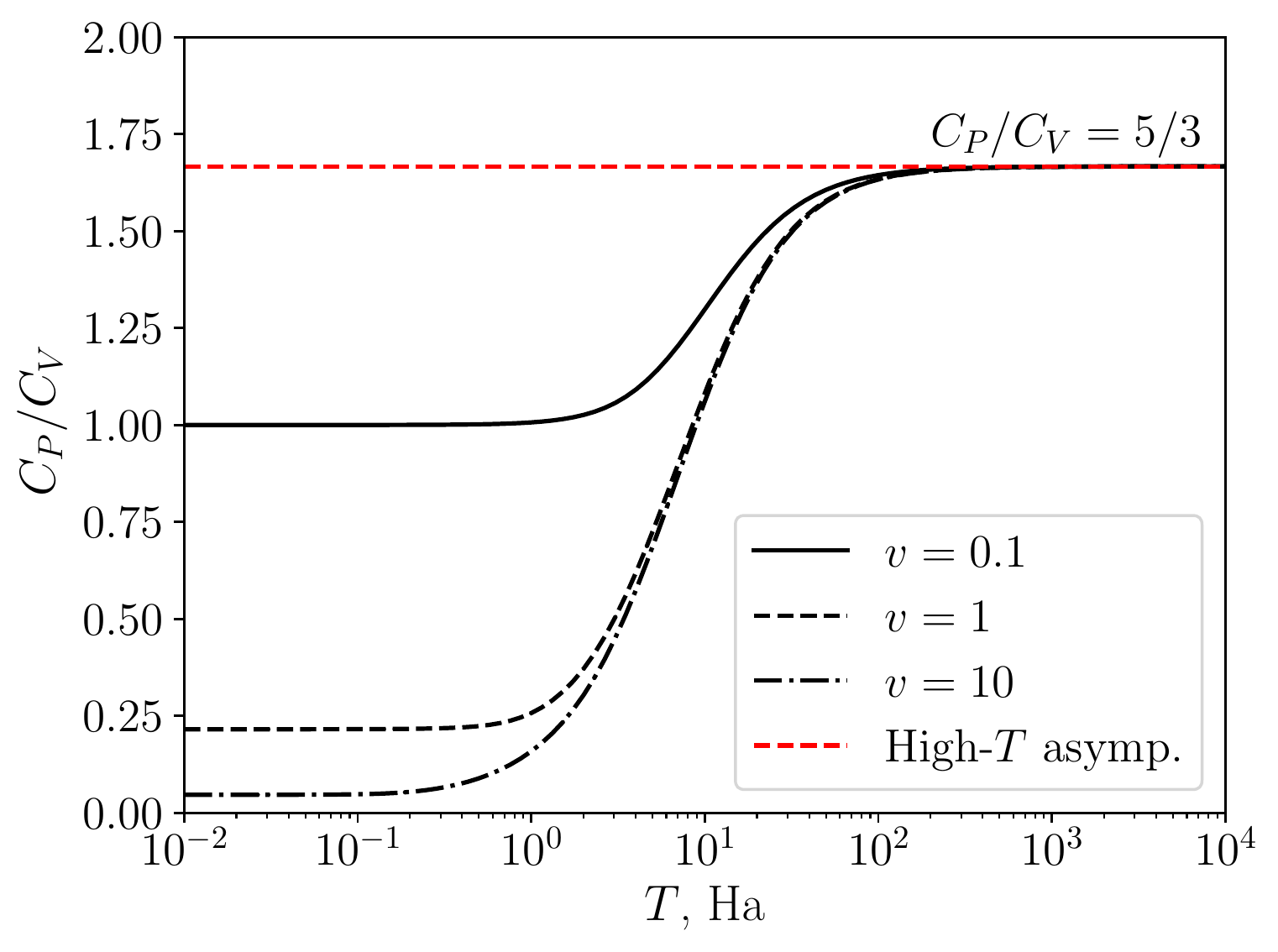}(a)
      \includegraphics[width=0.485\columnwidth]{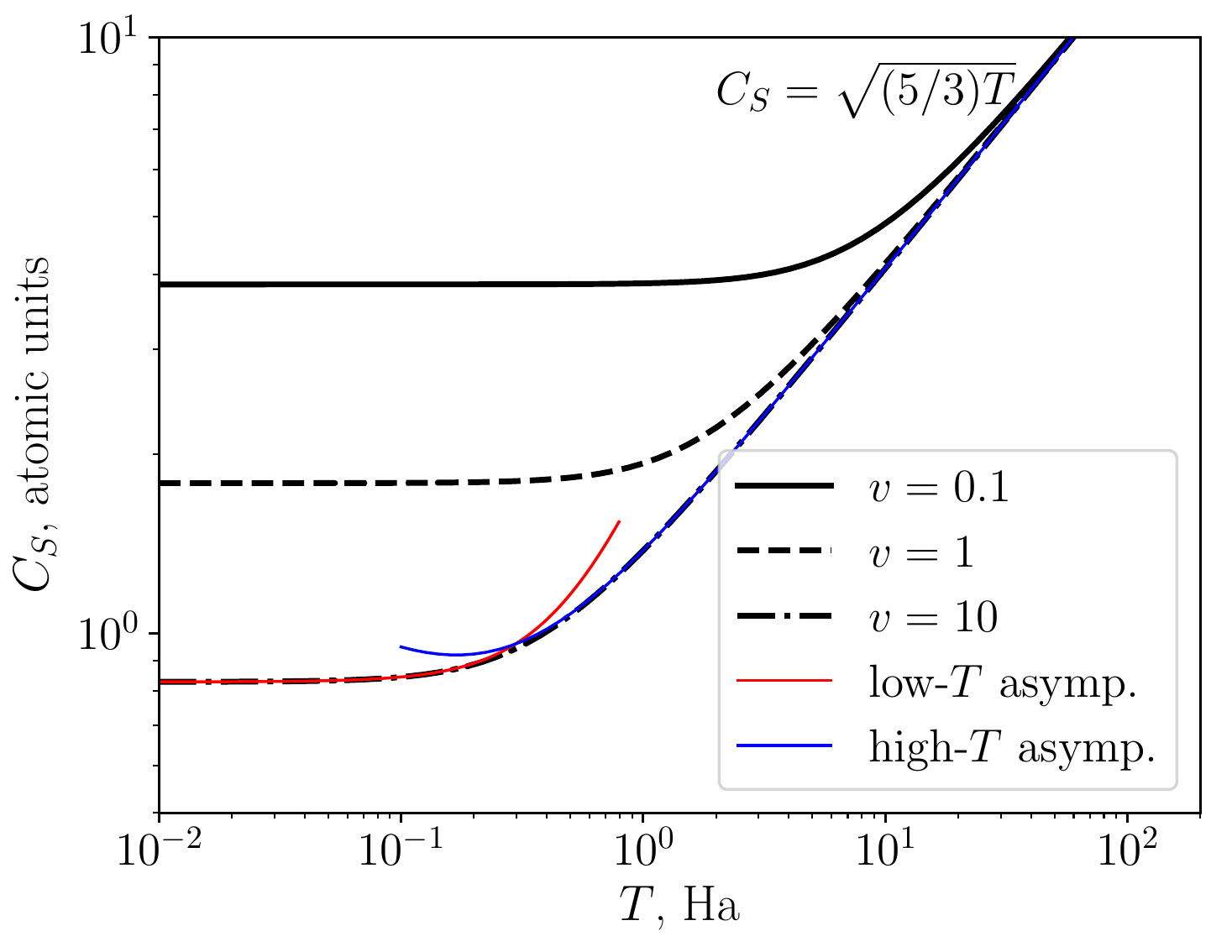}(b)
      \caption{Heat capacity ratio (a) and adiabatic sound velocity (b) of IFG at three isochores. Black lines: solid---$v = 0.1$, dashed---$v = 1$, dash-dotted---$v = 10$. Dashed red line---high-temperature asymptote for $C_P/C_V$; solid red line---low-temperature asymptote (\ref{eq:cs2_lt}) at $v = 10$; solid blue line---high-temperature asymptote (\ref{eq:CTCShighT}) at $v = 10$.}
      \label{fig:sound_velocity}
    \end{figure}

    Heat capacity ratio and adiabatic sound velocity can be seen in figure~\ref{fig:sound_velocity}. Again, we consider three isochores $v = 0.1$, 1 and 10. Heat capacity ratio $C_P / C_V$ is not constant for IFG and tends to a volume-dependent value at $T\to 0$. At high temperatures $C_P / C_V$ tends to a universal classical value $5/3$. Adiabatic sound velocity shows similar behavior; at high temperatures it tends to a volume-independent value $5/3T$.

    \subsection{Numerical implementation}
    Formulae (\ref{eq:pressure})--(\ref{eq:C_S}) provide a straightforward way to calculate IFG properties using \texttt{fdint} Python module \cite{fdint} for the Fermi--Dirac functions (\ref{eq:fermi-dirak_integral_definition}) calculation.
    However, formulae (\ref{eq:entropy}), (\ref{eq:C_v}) and (\ref{eq:C_P}) contain subtraction and thus may significantly suffer from the loss of accuracy. In this case the exact expressions should be replaced by the asymptotic ones in the low- and high-temperature limits. 

    Thermodynamic properties of IFG are implemented as a Python module and hosted on PyPi under the name \texttt{ifg} \cite{ifg}.
    The module supports calculation of energy, chemical potential, pressure, entropy, isobaric and isochoric heat capacities, isentropic and isothermal sound speeds. It is possible to use different system units and perform calculations for a given range of temperatures and volumes; output data can be easily represented in graphs.

    Due to the mentioned precision limitations, the following ranges of input parameters in atomic units are supported: $T \in [10^{-49},\, 10^{49}]$ and $v \in [10^{-30},\, 10^{20}]$. Inside these range 6-digit precision is guaranteed (tests are made with the Hypothesis \cite{MacIver2019Hypothesis} framework). The indicated ranges are by no means restrictive due to relativistic effects and the requirement of  thermodynamic equilibrium. 

    \section{Conclusions}
    Explicit analytical formulae for thermodynamic functions of non-relativistic ideal Fermi gas have been obtained. All functions including thermodynamic coefficients have been expressed through the first and second derivatives of the Helmholtz free energy. Asymptotic formulae at $T\to 0$ and $T\to\infty$ together with the first corrections have been also derived. Numerical calculation of thermodynamic functions has been implemented as a Python module available on PyPi \cite{ifg}. 
    

\section*{Acknowledgments}
The authors thank the Russian Science Foundation (Grant No.\,20-42-04421) for financial support.

%
%
\section*{References}


\providecommand{\newblock}{}

\nocite{*}
\end{document}